\documentclass{aa}  
\usepackage{keyval}
\usepackage{graphicx}
\usepackage{txfonts}
\usepackage{booktabs}
\usepackage{pdflscape}
\usepackage{longtable}  
\usepackage{placeins}
\usepackage{float}
\usepackage[colorlinks=False, linkcolor=blue]{hyperref}

\begin{document}

   \title{The long-term stability of the Vast Polar Structure and its connection to a possible previous passage of the LMC}

   \author{Alberto Manuel Martínez-García
          \and
          Andrés del Pino
          }

   \institute{Instituto de Astrofísica de Andalucía, CSIC, Glorieta de la Astronomía, 18080 Granada, Spain\\
              \email{ammartinez@iaa.csic.es}
             }

   \date{Received 24 July 2025; Accepted 4 September 2025}
   \titlerunning{The long-term stability of the VPOS} 
   \authorrunning{A. M. Martínez-García and A. del Pino}

  \abstract  
{
The Vast Polar Structure (VPOS) is a thin, planar arrangement of co-orbiting dwarf galaxies, nearly perpendicular to the Milky Way (MW) disc, whose origin and long-term evolution remain open questions.
}
{
In this work we investigate the persistence and stability of the VPOS over time.
}
{
We identify VPOS member galaxies and integrate their orbits over the past 5 Gyr using accurate phase-space data and time-evolving gravitational potentials that account for the mutual interaction between the MW and the Large Magellanic Cloud (LMC).
}
{
We identify 15 galaxies as members of the VPOS, including 9 MW and 6 LMC satellites. We find that the VPOS has remained a stable structure, maintaining a roughly constant thickness ($\sim$15 kpc), flattening ($c/a\sim 0.2$), and orientation over time.  While the LMC exerts a strong gravitational influence on the MW satellites, its impact on the VPOS is limited, leading only to mild perturbations. The structural properties of the VPOS remain almost unchanged, whether or not LMC satellites are included in the analysis, indicating a smooth dynamical integration with the rest of VPOS members upon entering the MW virial radius. This minimal dynamical impact on the VPOS results from the remarkable alignment between the LMC’s orbit and the plane’s orientation.
}
{
The VPOS is not a transient alignment, but a long-lived planar structure in the MW system, that has persisted for at least the last 5 Gyrs. Notably, the VPOS predates the LMC’s infall, ruling out formation scenarios tied to its recent approach to the MW. Our findings suggest a strong connection between the VPOS and the LMC, consistent with a scenario in which the LMC is on its second pericentre and the VPOS originated primarily from satellites stripped during the first passage.
}

   \keywords{galaxies: dwarf --
                 galaxies: kinematics and dynamics --
                galaxies: Magellanic Clouds --
                Local Group
               }

   \maketitle

\section{Introduction}
The Milky Way (MW) is currently thought to host at least 60 satellite galaxies \citep{Doliva-Dolinsky2025}. The most prominent, and earliest known among them are the Magellanic Clouds,  easily visible to the naked eye in the southern hemisphere sky, and familiar to its inhabitants for millennia.  It was not until 1938 that the next satellite, the Sculptor dwarf spheroidal galaxy, was discovered \citep{Shapley1938}, followed shortly thereafter by Fornax \citep{Shapley1938b}.  Since then, and especially with the advent of CCD technology and wide-field surveys (e.g. the Sloan Digital Sky Survey; \citealt{York2000}), the census of MW satellites has grown significantly. This population is very likely to continue growing in the near future, just with the Vera Rubin Large Survey of Space and Time (LSST) being expected to find more than eighty new  satellites (\citealt{Tsiane2025}).

In the early stages of the study of the MW satellites, it was already noted that some of the dwarfs known at the time, in particular Draco, Sculptor and Ursa Minor, 
plus the Magellanic Stream and some globular clusters lied in a great circle structure perpendicular to the MW disc 
(\citealt{Kunkel1976, Lynden-Bell1976}). 
The presence of that alignment was repeatedly confirmed by the successive discoveries of new satellites following the same planar distribution (e.g. \citealt{Kroupa2005, Metz2007}). 
Known as the Vast Polar Structure (VPOS, \citealt{Pawlowski2012, Pawlowski2013}),  such arrangement of satellites displays a thickness of 20-30 kpc, short-to-long axis ratio of $0.18-0.30$ and 
radius of $\sim 250$ kpc (\citealt{Pawlowski2018, Pawlowski2021}), with the majority of its members showing a coherent motion (\citealt{Metz2008, Fritz2018, Li2021, Taibi2024}). 
The VPOS is not a unique feature of the MW system. Other planes of satellites  have also been reported around M31 (\citealt{Metz2007,Ibata2013, Conn2013}), and in galaxies beyond of the Local Group (LG), like M81 (\citealt{Chiboucas2013}), M101 (\citealt{Merritt2014}), Centaurus A (\citealt{Muller2018}),  NGC 253 (\citealt{MartinezDelgado2021}), and NGC 2750 (\citealt{Paudel2021}), among others (see \citealt{Pawlowski2021}).

Planes of satellites, and most notably the VPOS, have been a significant subject of debate during the last two decades. It has been argued that the likelihood of the existence of such  arrangements is very low in the context of the currently most widely accepted cosmological framework, the $\Lambda$ Cold Dark Matter ($\Lambda$CDM) model. \citet{Kroupa2005} showed that the planar distribution of the 11 classical satellites of the MW is very unlikely should it had been drawn from an isotropic distribution. Many studies based on cosmological simulations have searched for similar anisotropic phase-space configurations, consistently finding that they are very rare (see \citealt{Pawlowski2018}).
A recent example is the case of the plane of satellites  observed around NGC 4490, for which comparable structures were found in only 0.21\% of analogous galaxies in the TNG50 cosmological simulation \citep{Karachentsev2024, Pawlowski2024}, highlighting how uncommon such configurations are in the $\Lambda$CDM model.
 This mismatch between predictions and observations is often referred to as the \textit{Planes of Satellite Galaxies Problem} \citep{Pawlowski2018}. However, it is debated whether it is another possible $\Lambda$CDM small-scale 
problem  or if it does not challenge it at all (see \citealt{Bullock2017}). 

The formation of the VPOS, and of planes of satellites in general, remains another subject of ongoing debate. Several mechanisms have been proposed to explain these anisotropic distributions of satellites, however there is not yet a clear, definitive solution to the problem (\citealt{Pawlowski2018}). Some possible explanations are the preferential accretion of matter through filaments (\citealt{Libeskind2011}) or the accretion of satellites in groups (\citealt{LiHelmi2008, Smith2016}). It has also been  proposed that the members of the VPOS could be tidal dwarf galaxies, formed from the debris resulting of an encounter between the early MW and another galaxy (\citealt{Pawlowski2011}). In recent years, several scenarios have been proposed in which the Large Magellanic Cloud (LMC) plays a significant role in the formation of the VPOS. One scenario proposes  that the recent close passage of the LMC could have contributed to the clustering of the orbital poles of the MW satellites (\citealt{GaravitoCamargo2021}). Another suggests that if the LMC is in its second pericentric passage around the MW, the VPOS could naturally arise from the stripping of some of its satellites during the first pericentre (\citealt{Vasiliev2024}).

In order to better understand the origin of the VPOS and  to assess the degree to which it challenges the $\Lambda$CDM model, it is essential to study its stability and persistence in time.
Several works have tried to tackle this problem by integrating  the orbits of the galaxies of the  VPOS  and then tracking the evolution of its thickness and axis ratios over time. \citet{Lipnicky2017} made use of proper motions (PM) from the \textit{Hubble Space Telescope} (HST) for the classical MW satellites and integrated their orbits in a static MW potential, finding that the significance of the plane vanished within a mere 1 Gyr. Similarly, \citet{Maji2017} performed forward integration of the orbits of the same 11 satellites to predict the future of the VPOS. 
They concluded that the VPOS would triple its width in 1 Gyr, and that ultimately the galaxies would abandon the plane.
More recently, \citet{Sawala2023} has also argued that the plane is a transient structure, based on forward and backward integration  using systemic PMs derived with \textit{Gaia} Early Data Release 3 (EDR3, \nocite{GaiaEDR3}Gaia Collaboration 2021). They found that the plane gets thicker both towards past and future times, and also that the orientation of the plane has changed $\sim 17$ degrees over the last 0.5 Gyrs.

All these studies agree that the VPOS is a short-lived, unstable structure. However, it remains unclear to what extent the observed thickening of the plane over time reflects a true physical dispersion, or whether it arises from the propagation of observational uncertainties during the orbit integration (\citealt{Kumar2025}).
Moreover, previous analyses exclusively assumed static, MW-only  potentials for the orbit integration. In contrast, over the last years, multiple studies based on observational data and cosmological simulations have shown that the gravitational influence of the LMC significantly affects the orbits of MW satellites (e.g., \citealt{Patel2020, Pace2022, Battaglia2022, Souza2022}; see \citealt{Vasiliev2023} for a review). This impact is non-negligible and thus must be taken into account when studying the stability of the VPOS.

In this paper, we present a new analysis of the VPOS’s long-term stability based on orbit integrations of its present-day members. Using accurate 6D phase-space coordinates as initial conditions, we compute the VPOS member galaxies trajectories under six time-evolving gravitational potentials that include the mutual interaction between the MW and the LMC. This approach allows us to trace the evolution of the VPOS’s thickness, shape, and orientation over the past 5 Gyr, providing a direct assessment of its persistence and dynamical stability.
The paper is organized as follows. In Section~\ref{sec:methods} we introduce the data and methods, in Section~\ref{sec:resuts} we present and discuss the results and finally in Section~\ref{sec:conclusions} we summarize and present the main conclusions of this work.

\section{Methods}
\label{sec:methods}
To investigate the stability of the VPOS over time, we first need to identify which galaxies are currently part of it in order to backward integrate their orbits. The reconstructed orbits provide us with a record of the positions of these galaxies as a function of time and thus allow us to trace the evolution of the VPOS. In this Section we introduce the sample of galaxies to be studied, their phase-space coordinates, and the procedure used for the VPOS membership classification. We then outline the gravitational potentials used for the orbit integration, the integration procedure itself, and the plane fitting method used to quantify the structure’s stability and orientation over time.  We note that we only present a brief overview of the galaxy sample, the potentials and the integration method, since full details will be available in Martínez-García et al. (in prep.), a dedicated paper on the orbit integration of dwarf satellites around the MW.

\subsection{Galaxy sample}
Our study is based on a sample of 58 dwarf galaxies located within 500 kpc of the MW, for which 6D phase-space coordinates are available. The core of the sample is drawn from \citet{Battaglia2022}, which provides accurate systemic PMs for a large number of LG dwarfs based on \textit{Gaia} EDR3, alongside a compilation sky coordinates, distances, and radial velocities from the literature. We complement this core sample with additional  systems, namely Aquarius III, Bootes V, Eridanus IV, Leo VI, Pegasus IV, the Sagittarius dSph, and the Small Magellanic Cloud (SMC), whose  phase-space coordinates were obtained from the \textit{Local Volume Database} \citep{Pace2024}. We note that the systemic PMs of the galaxies of the sample are primarily based on Gaia EDR3/DR3, except for the SMC, for which we adopt values from HST \citep{Zivick2018}. To ensure accurate orbit integration, we accounted for the effect of \textit{Gaia} EDR3/DR3  systematics in the PMs using a procedure analogous to the one described in \citet{Battaglia2022}, with the exception that we made use of the newly available QSO candidate table released with Gaia DR3 (\texttt{qso\_candidate}, \nocite{GaiaDR3QSO} Gaia Collaboration 2023).

\subsection{VPOS membership selection}
\label{sec:VPOSmem}
We determine which galaxies of the sample are members of the VPOS, following a procedure analogous to that of \citet{Taibi2024}. First, we compute the specific angular momentum of each galaxy. To account for observational uncertainties, we perform a Monte Carlo (MC) scheme with $10^3$ iterations, that randomly samples the phase-space coordinates from normal distributions centered on their measured values, with standard deviations equal to their reported uncertainties. When uncertainties are asymmetric, we take as standard deviation their mean value. The sampled positions and velocities
are then transformed into Galactocentric Cartesian coordinates with Astropy  (\nocite{Astropy2013, Astropy2018}Astropy Collaboration 2013, 2018) using the following solar parameters: $R_{0} = 8.122 \pm 0.021$ kpc (distance from the Sun to the Galactic center; \nocite{GRAVITY2018} GRAVITY Collaboration 2018), $z_{\odot} = 20.8 \pm 0.3$ pc (height of the Sun above the Galactic midplane; \citealt{BennettBovy2019}), $V_{R, \odot} = -12.9 \pm 3.0$ km s$^{-1}$ (radial velocity of the Sun with respect to the Galactic center; positive values are directed outward from the center), $V_{\phi, \odot} = 245.6 \pm 1.4$ km s$^{-1}$ (tangential velocity of the Sun in the direction of Galactic rotation), and $V_{Z, \odot} = 7.78 \pm 0.09$ km s$^{-1}$ (vertical velocity of the Sun relative to the Galactic midplane, directed toward the North Galactic Pole; \citealt{DrimmelPoggio2018}). These parameters are also randomly sampled in each iteration. 
For each of these $10^3$ samples, we compute the orbital poles, i.e. the direction of the specific angular momentum vector, via the cross product of position and velocity, and  transform them into Galactic coordinates. 
We then quantify the alignment of the orbital poles with the VPOS normal vector, which points towards $(l,b) = (169.3^\circ, -2.8^\circ)$ \citep{PawlowskiKroupa2013, Fritz2018}. For each galaxy, we compute the fraction of MC realizations whose orbital poles fall within the region defined by the area covering 10\% of the sky around the VPOS normal vector direction (with an aperture of $36.87^\circ$), and its antipole. 
This fraction, $f_{\mathrm{VPOS}}$, is as a measure of the alignment with the VPOS. Galaxies with $f_{\mathrm{VPOS}} > 0.95$ are classified as 'on-plane', those with $f_{\mathrm{VPOS}} < 0.05$ as 'off-plane', and systems in the range $0.05 < f_{\mathrm{VPOS}} < 0.95$ are considered 'uncertain'.

\subsection{Orbit integration}
\subsubsection{Gravitational potentials}
We integrate the orbits of the VPOS on-plane galaxies using six time-evolving gravitational potentials that account for the mutual interaction between the MW and LMC, as modeled by N-body simulations of their dark matter (DM) halos (see \citealt{Vasiliev2023, Vasiliev2024}).

The first potential ("V23", \citealt{Vasiliev2023}) consists of a MW model with an exponential disk of total mass $5 \times 10^{10}$ M$_{\odot}$, a spherical bulge of $1.2 \times 10^{10}$ M$_{\odot}$ and a triaxial DM halo of $8 \times 10^{11}$ M$_{\odot}$.  The LMC is modeled by a truncated NFW profile with initial mass $1.5 \times 10^{11}$ M$_{\odot}$, scale radius of 10.84 kpc, truncation radius of 108.4 kpc, and its trajectory around the MW has a single pericentre.  This model is the same as the evolving triaxial MW + $1.5 \times 10^{11}$ M$_{\odot}$ LMC potential of \citet{Vasiliev2021}, that successfully fits the Sagittarius stream, with the exception that allows to explore longer integration times.

The rest of potentials are combinations of the MW and LMC models from \citet{Vasiliev2024}, in which the LMC trajectory has two pericentres. The DM halos of both galaxies consist of  truncated NFW profiles. The LMC initial virial masses are $1.92 \times 10^{11}$ M${\odot}$ ("L2") and $2.76 \times 10^{11}$ M${\odot}$ ("L3") respectively, and decrease with time due to the interaction with the MW, mostly during the LMC's earliest pericentre. The MW models have initial virial masses of $10.0 \times 10^{11}$ M${\odot}$ ("M10") and $11.0 \times 10^{11}$ M${\odot}$ ("M11"), and their mass profiles match the reported observational constrains at present time (see \citealt{Wang2020} and Figure 1 of \citealt{Vasiliev2024}). Both MW models share the same static baryonic disk ($5 \times 10^{10}$ M${\odot}$) and bulge ($1.2 \times 10^{10}$ M${\odot}$). The four resulting combinations of MW-LMC models (e.g. L2M10) allow us to explore a wide range of realistic configurations for the MW and LMC system. Lastly we include a variant of the model L2M10 with a single passage ("L2M10first").

\subsubsection{Integration procedure}
For orbit integration, we use the AGAMA software \citep{AGAMA2018, AGAMA2019} embedded  within a MC scheme  in order to account for the observational uncertainties. 
For each galaxy and potential, we run $10^3$ iterations, sampling phase-space coordinates from random normal distributions centered
on their measured values and dispersion equal to their uncertainties.
The sampled coordinates are then transformed into Galactocentric Cartesian coordinates. With these initial conditions,  orbits are integrated backward using AGAMA’s \texttt{orbit} method in 1500 steps from  $t = 0$ to $t = -5$ Gyr (or $-4$ Gyr for the L2M10first potential\footnote{Potential L2M10first only allows us to integrate over the last 4 Gyrs, since the simulated time for such potential does not reach earlier times (see~\citealt{Vasiliev2024}).}). We note that the effect of dynamical friction is not considered in the integration, however we do not expect it to have a significant impact in our analysis (we address the reader to Appendix~\ref{sec:app2}, where we further discuss this and other possible caveats of our work). We limit the orbit integration to the past 5 Gyr, as MW-like halos are typically assembled by that epoch, with no major mergers occurring afterward \citep{Santistevan2020, Sotillo-Ramos2022}. Going further back in time would enter a regime where significant accretion events or massive mergers not captured by our time-evolving potentials are likely to affect the satellites' dynamics, making the integration results increasingly unreliable. The output of the integration  consists of a set of $10^3$ orbits per galaxy and potential.

\subsection{Plane stability}
Once the orbits of the  member galaxies of the VPOS have been derived, we can study its evolution. To assess both the persistence, and stability  of the VPOS for the  different gravitational potentials, we adopt a recursive approach
consisting of $10^3$ iterations, each of them performing the following procedure. 
For a given potential, we randomly select, for each on-plane galaxy, one orbit from its corresponding set of MC realizations.
From the chosen orbits, we examine the positions of the galaxies at all times to evaluate the planarity and orientation of the VPOS. For each integration time
we perform a Principal Component Analysis (PCA) on the 3D positions of the galaxies. In this process, we first calculate the covariance matrix of the galaxy spatial distribution (analogous to the inertia tensor) and then derive its eigenvectors, which represent the principal axes of the galaxy distribution, and  eigenvalues, which indicate the variance along these axes.
From the eigenvalues, we determine the axis ratios $c/a$ and $b/a$. The $c/a$  ratio, represents the short-to-long axis ratio and measures how thin the galaxy distribution is. A low $c/a$ value implies a thin, flattened structure. The $b/a$ ratio, representing the intermediate-to-long axis ratio, provides insight into the shape of the distribution. If $b/a$ is close to 1, the distribution is more disc-like or oblate, indicating a relatively extended structure in two dimensions. Conversely, if $b/a$ is small and similar to $c/a$, the distribution is more elongated or prolate.
We then measure the thickness of the plane by projecting the galaxies' positions along the direction of its normal vector, i.e. the shortest principal axis of the galaxy distribution. We calculate the root mean square (RMS) of these projected distances, which quantifies the spread or thickness of the galaxy distribution with respect to the plane. Additionally, we calculate the Median Absolute Deviation (MAD) of these distances, which offers robustness against extreme values, providing a more resilient measure of thickness.
We also determine the orientation of the VPOS from the normal vector of the plane.

This procedure is repeated in each iteration, yielding as output a set of $10^3$ measurements of plane thickness, axis ratios, and normal vector, for each integration time. To characterize the evolution of these quantities, we compute  at each time step their median values across iterations. Uncertainties are estimated from the 16th and 84th percentiles of the distributions.
For the normal vectors, we also compute the mean resultant length (MRL) across iterations to quantify the directional coherence as a function of time. The MRL ranges from 1 (perfect alignment between vectors) to 0 (complete random distribution), serving as a standard metric for assessing concentration in directional data. However, the normal vector of a plane is inherently ambiguous up to a sign, a vector and its opposite represent the same plane. To solve this ambiguity, we adopt a consistent convention: we select a fix reference direction and, for each normal vector, enforce its alignment by flipping it if it forms an angle greater than 90° with the reference. This effectively confines the normal vectors to a hemisphere, such that the MRL in our analysis ranges from 1 (perfect alignment) to 0.5 (uniform distribution over a hemisphere).

\section{Results and discussion}
\label{sec:resuts}

\subsection{Members of the VPOS}
\label{sec:members}

\begin{figure*}
    \centering
    \includegraphics[width=0.9\linewidth]{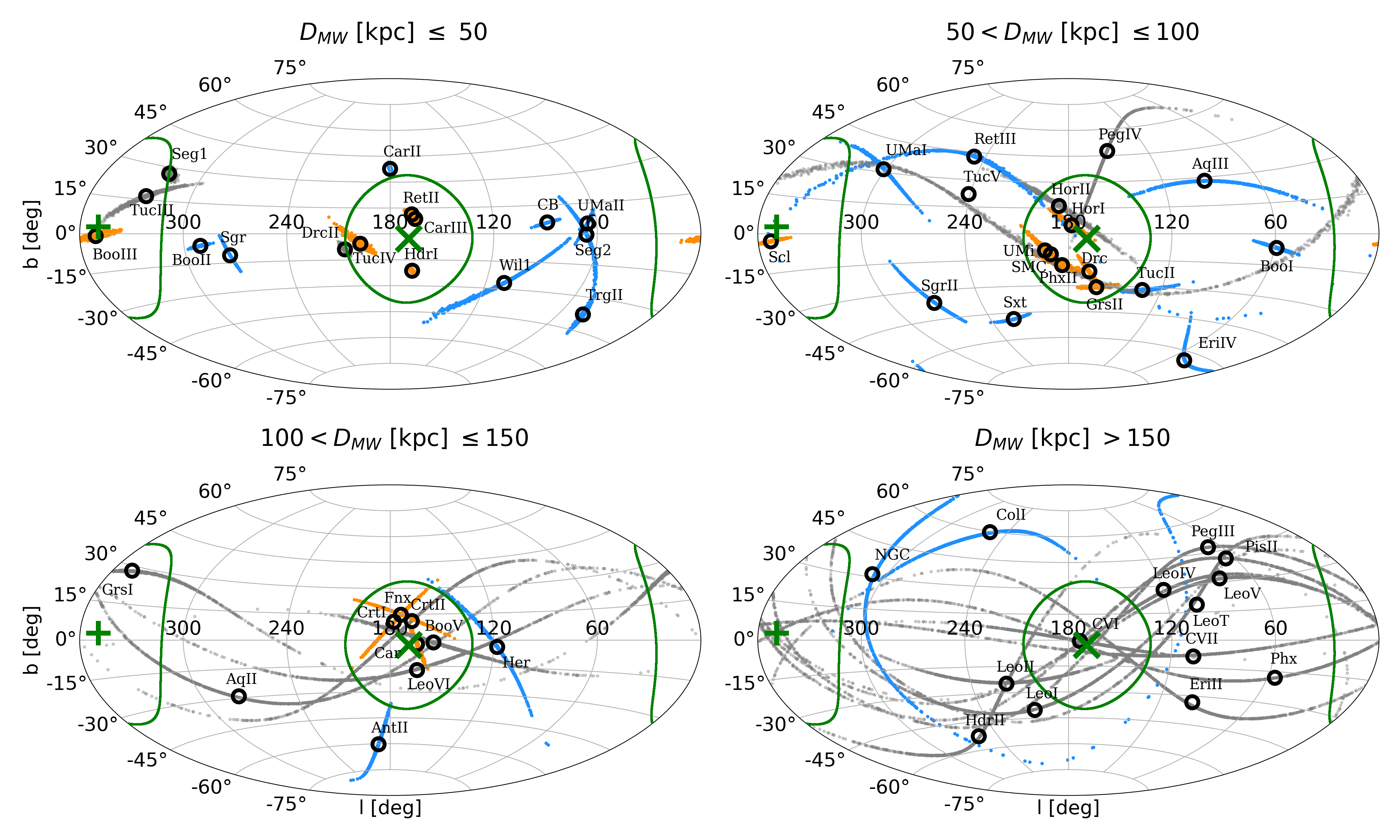}
    \caption{Distribution of the orbital poles for 58 dwarf galaxies around the MW in Galactic coordinates. Solid points represent the orbital poles from the MC realizations of each galaxy. The color of the points indicates VPOS membership: orange for on-plane galaxies, blue for off-plane, and grey for uncertain membership. Empty circles mark the median orbital pole of each galaxy. Green crosses denote the VPOS normal vector direction as reported in \citet{PawlowskiKroupa2013}, green '+' marker denotes the opposite direction, while green lines outline the region covering 10\% of the sky around them. Labels are abbreviated versions of the galaxy names. Each plot shows only the poles of galaxies within a specific Galactocentric distance range to ease visualization.
    }
    \label{fig:VPOSt0}
\end{figure*}

We studied the potential membership of 58 dwarf galaxies to the VPOS by analyzing the distribution of their orbital poles. 
The full list of studied galaxies, along with their membership classification, and $f_{\mathrm{VPOS}}$ are presented in Table~\ref{tab:vpos_membership}. 
The distribution of the orbital poles in Galactic coordinates is shown in Figure~\ref{fig:VPOSt0}. For each galaxy we represent the orbital poles from the MC realizations and their median values. Galaxies are marked according to their membership to the VPOS and they are split across panels based on their Galactocentric distance to ease visualization.

\begin{table}[ht]
    \caption{VPOS membership classification}
    \centering
    \fontsize{8}{10}\selectfont
    \setlength{\tabcolsep}{4pt} 
    \begin{tabular}{lcl  lcl}
    \toprule
    \toprule
    Galaxy & Class. & $f_{\mathrm{VPOS}}$ & Galaxy & Class. & $f_{\mathrm{VPOS}}$ \\
    \cmidrule(lr){1-3} \cmidrule(lr){4-6} \cmidrule(lr){1-3} \cmidrule(lr){4-6}
Antlia II & off & 0.00 & Leo II & ? & 0.25 \\
Aquarius II & ? & 0.28 & Leo IV & ? & 0.32 \\
Aquarius III & off & 0.00 & Leo T & ? & 0.17 \\
Bootes I & off & 0.00 & Leo V & ? & 0.14\\
Bootes II & off & 0.00 & Leo VI & ? & 0.58\\
Bootes III & on & 1.00 & NGC 6822 & off & 0.00\\
Bootes V & ? & 0.71 & Pegasus III & ? & 0.38 \\
C. Venatici I & ? & 0.80 & Pegasus IV & ? & 0.23\\
C. Venatici II & ? & 0.38 & Phoenix & ? & 0.37\\
Carina & on & 1.00 & Phoenix II & on & 1.00\\
Carina II & off & 0.00 & Pisces II & ? & 0.36 \\
Carina III & on & 1.00 & Reticulum II & on & 1.00\\
Columba I & off & 0.00 & Reticulum III & off & 0.04 \\
C. Berenices & off & 0.00 & SMC & on & 1.00 \\
Crater I & ? & 0.58 & Sagittarius & off & 0.00 \\
Crater II & on & 1.00 & Sagittarius II & off & 0.00 \\
Draco & on & 1.00 & Sculptor & on & 1.00 \\
Draco II & ? & 0.27 & Segue 1 & ? & 0.70 \\
Eridanus II & ? & 0.30 & Segue 2 & off & 0.00 \\
Eridanus IV & off & 0.00 & Sextans & off & 0.00 \\
Fornax & on & 0.99 & Triangulum II & off & 0.00 \\
Grus I & ? & 0.64 & Tucana II & off & 0.00 \\
Grus II & on & 1.00 & Tucana III & ? & 0.85 \\
Hercules & off & 0.00 & Tucana IV & on & 0.98 \\
Horologium I & on & 1.00 & Tucana V & ? & 0.37 \\
Horologium II & ? & 0.76 & Ursa Major I & off & 0.00 \\
Hydra II & ? & 0.09 & Ursa Major II & off & 0.00 \\
Hydrus I & on & 1.00 & Ursa Minor & on & 1.00\\
Leo I & ? & 0.35 & Willman 1 & off & 0.00 \\
\bottomrule
    \end{tabular}
    \tablefoot{Studied galaxies and their membership to the VPOS. Column 1 shows the names of the galaxies, Column 2 their classification as on-plane (on), off-plane (off) or uncertain (?) members of the VPOS, and Column 3 the fraction of MC realizations where there orbital poles fall within the area encompassing the 10\% of the sky around the VPOS normal direction and its antipole. Columns 4-6 follow the same structure.}
    \label{tab:vpos_membership}
\end{table}

We find 15 on-plane galaxies. They are a combination of 5 classic MW satellites (Carina, Draco, Fornax, Sculptor and Ursa Minor), 4 ultra faint dwarf (UFD) MW satellites (Bootes III, Crater II, Grus II, Tucana IV)  and 6 potential LMC satellites (Carina III, Horologium I, Hydrus I, Phoenix II, Reticulum II and the SMC, see \citealt{Vasiliev2024} for the probability of association of each galaxy with the LMC). All the galaxies are co-rotating in the plane, except for Bootes III and Sculptor, that are counter-rotating and thus have their orbital poles in the antipode of the VPOS area (see Fig.~\ref{fig:VPOSt0}). We note that in our study we do not consider the LMC for the VPOS membership classification since for the orbit integration we assume its trajectory has no uncertainties. However, it is a well known member of the VPOS (e.g. \citealt{Fritz2018, Taibi2024}). 
The off-plane population consists of 20 dwarfs. They are mostly UFDs but we also find two of the  classic satellites of the MW (Sextans and Sagittarius).
Finally, the majority of the studied galaxies, 23 out of 58, show an uncertain degree of association with the VPOS. These systems typically show a large spread in their orbital pole distributions, not allowing us to draw strong conclusions about their alignment with the VPOS. Most of them are UFD, for which current measurements of PMs and distances are usually less precise.

We note that most of the likely LMC satellites in our sample are classified as on-plane, reflecting their compact configuration and shared motion with the LMC system as it approaches the MW. However, two notable exceptions are Carina II and Horologium II. Despite being formally classified as off-plane and uncertain, respectively, both exhibit kinematics consistent with the LMC system and thus are potentially aligned with the VPOS. Carina II has well-constrained phase-space coordinates and tightly clustered orbital poles that lie just outside the VPOS-defined region (see Fig.~\ref{fig:VPOSt0}), suggesting a likely membership to the VPOS. Horologium II displays a broader pole distribution, but a majority of its realizations still fall within the VPOS region, indicating a plausible, though less certain, association. These cases underscore the difficulty in assigning sharp boundaries to VPOS membership and the importance of considering individual orbital histories alongside formal classifications.

\begin{figure*}
    \centering
    \includegraphics[width=0.75\linewidth]{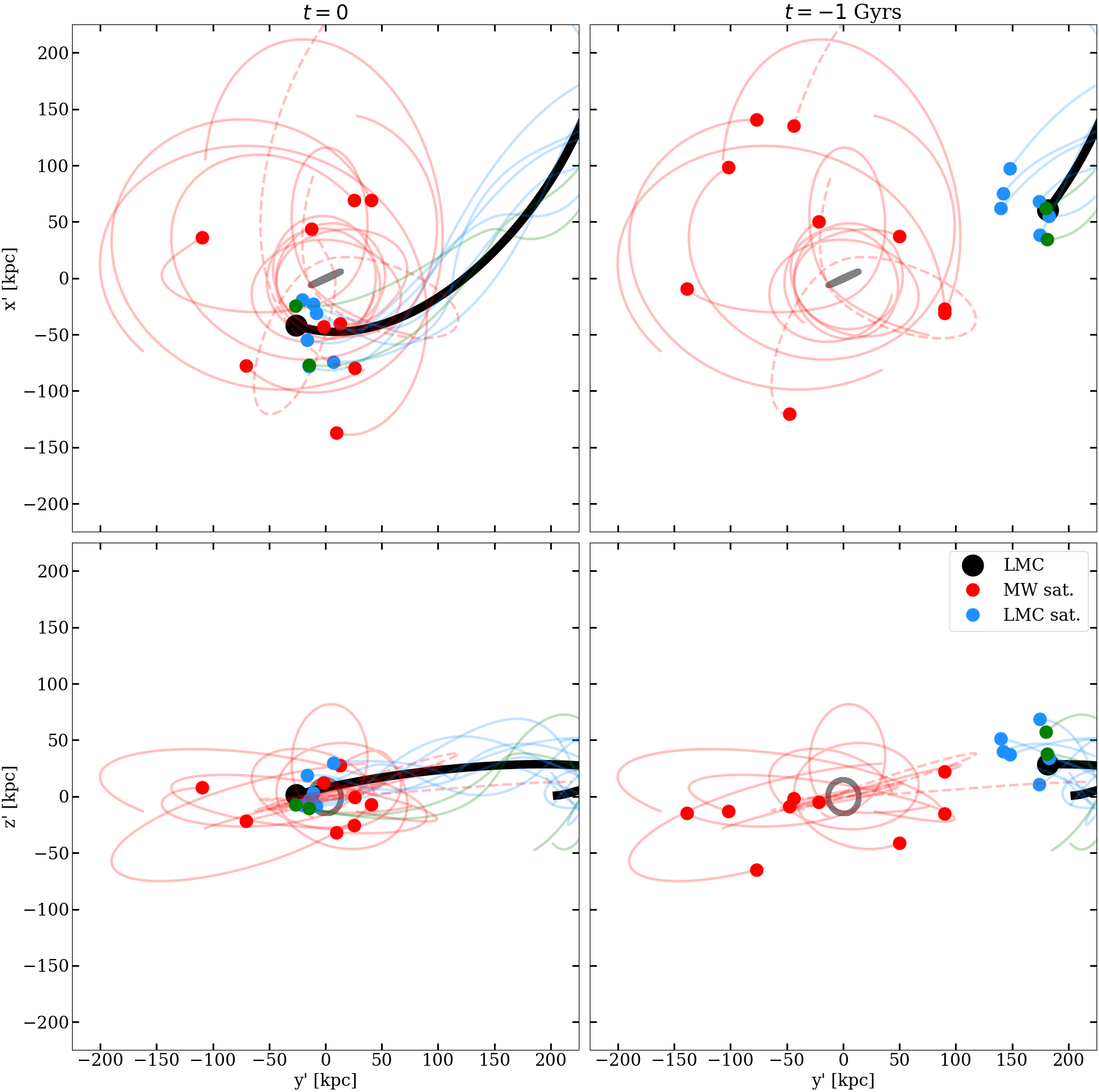}
    \caption{Spatial distribution and orbital histories of VPOS member galaxies. The panels show the positions of on-plane galaxies and their past trajectories under potential V23, at the present time (left) and at $t = -1$ Gyr (right). The coordinate system used ($x', y', z'$) is not Galactocentric, but instead defined by the principal axes of the galaxy distribution at $t = 0$; in this frame, the $z'$-axis is aligned with the normal vector of the VPOS. Points indicate galaxy positions at the corresponding time, while lines trace their past trajectories. Red denotes MW satellites, blue indicates LMC satellites, and black marks the LMC itself. Carina II and Horologium II are shown in green. Although not formally classified as on-plane members, they are LMC satellites and thus likely part of the VPOS (see Section~\ref{sec:members}). Dashed lines are used to represent the trajectories of Bootes III and Sculptor, the two counter-rotating galaxies of the VPOS. A grey annulus at the origin represents the MW stellar disk for reference, with a radius of 15 kpc.}
    \label{fig:plane_orbits}
\end{figure*}

To further illustrate the spatial and kinematic coherence of the on-plane galaxies of the VPOS, in Figure~\ref{fig:plane_orbits} we represent their orbits. The orbits shown were integrated under the potential V23 using the nominal phase-space coordinates of each galaxy, without incorporating uncertainties. The figure shows face-on and edge-on views of the VPOS at the present time and at $t = -1$ Gyr, a moment chosen to better
distinguish the MW and LMC satellite populations before they
further approach and become difficult to disentangle.
For context, Carina II and Horologium II are also shown but marked differently. The figure highlights the clear planar alignment and coherent motion of most on-plane satellites, with Bootes III and Sculptor standing out as notable counter-rotating outliers. Sculptor appears to come from large distances and has only recently completed a pericentre  ($t \sim -0.5$ Gyr), a behavior also observed in potentials L3M10 and L3M11. In the other potentials, however, it completes multiple pericentres, resembling Bootes III, which has three in the past 5 Gyr.
In parallel, the LMC and its associated satellite group clearly form a  compact, dynamically coherent system following a trajectory closely aligned with the VPOS plane. In the earlier snapshot ($t = -1$ Gyr), the LMC group is seen approaching an already-established planar structure around the MW, suggesting that the plane of satellites predates the LMC’s recent infall and was not formed as a result of it.

Our results regarding the VPOS members are very similar to the ones reported by \citet{Taibi2024}. This is due to the fact that both studies make use of a similar sample of galaxies and phase-space coordinates. The classification of galaxies as  on-plane, off-plane and uncertain VPOS members are in excellent agreement, with the only discrepancy being Tucana V. In our study it is a uncertain member of the VPOS (instead of off-plane), and shows a larger spread in the orbital poles, what is likely to be due to the large and asymmetric uncertainties in the PM of Tucana V (see \citealt{Battaglia2022}). \citet{Taibi2024} uses a bi-variate distribution when sampling the PMs in the MC, whereas we use the mean of the asymmetric uncertainties what is likely to produce a larger dispersion of the orbital poles when the asymmetry of the PM uncertainties is significant. 
We note that our work explores the VPOS membership for a larger number of galaxies, 58 in total. We study 9 more galaxies,  but we do not find any new VPOS members. We find that Aquarius III, Eridanus IV, Sagittarius, and NGC6822 are off-plane galaxies,  and that  the association of Bootes V, Leo VI, Leo T, PegasusIV, and Phoenix, with the VPOS is  unclear. 

Finally, we explored how VPOS membership may have evolved over time by recalculating the orbital poles of each galaxy  every 1 Gyr from their integrated orbits. As expected, the propagation of uncertainties in the integration cause the orbital poles to spread with time, making the $f_{\mathrm{VPOS}} > 0.95$ criterion increasingly strict at earlier epochs. Despite this, some galaxies (e.g., Bootes III, Tucana IV or Ursa Minor) still meet it even at $t = -5$ Gyr. Most other on-plane galaxies retain  their median orbital poles within or very near the VPOS area throughout the whole integration time. Some off-plane or uncertain members of the VPOS, like Draco II, Segue 1, and Tucana III, whose orbital poles lie in the border of the VPOS area or are partially inside at $t = 0$, may also align with the VPOS at earlier times, what highlights that using strict boundaries to define the membership can obviate some possible members. A striking case is Grus II, which usually stays well outside the VPOS area for most of the integration time and only enters around $t \sim -0.5$ Gyr, becoming an on-plane galaxy. The timing of the realignment of its orbital pole, together with the analysis of its orbital history, suggests that it was likely caused by the recent close passage of the LMC, which reoriented Grus II’s trajectory (Martínez-García et al., in prep.). We also investigated the converse scenario to the case of Grus II, namely, whether any galaxies that were originally consistent with the VPOS have been scattered out of it by the influence of the LMC, and found no evidence of such an event. We thus conclude that the on-plane population of the VPOS has remained stable over the past 5 Gyr, with the notable exception of the recent inclusion of Grus II.
To avoid potential bias in the study of the VPOS stability, we exclude Grus II from such analysis. Nonetheless, for clarity and consistency, we continue to use the term “VPOS members” or “on-plane galaxies” to refer to the remaining satellites.

\subsection{The VPOS over time}
\label{sec:vposevo}

\begin{figure*}
    \centering
    \includegraphics[width=0.95\linewidth]{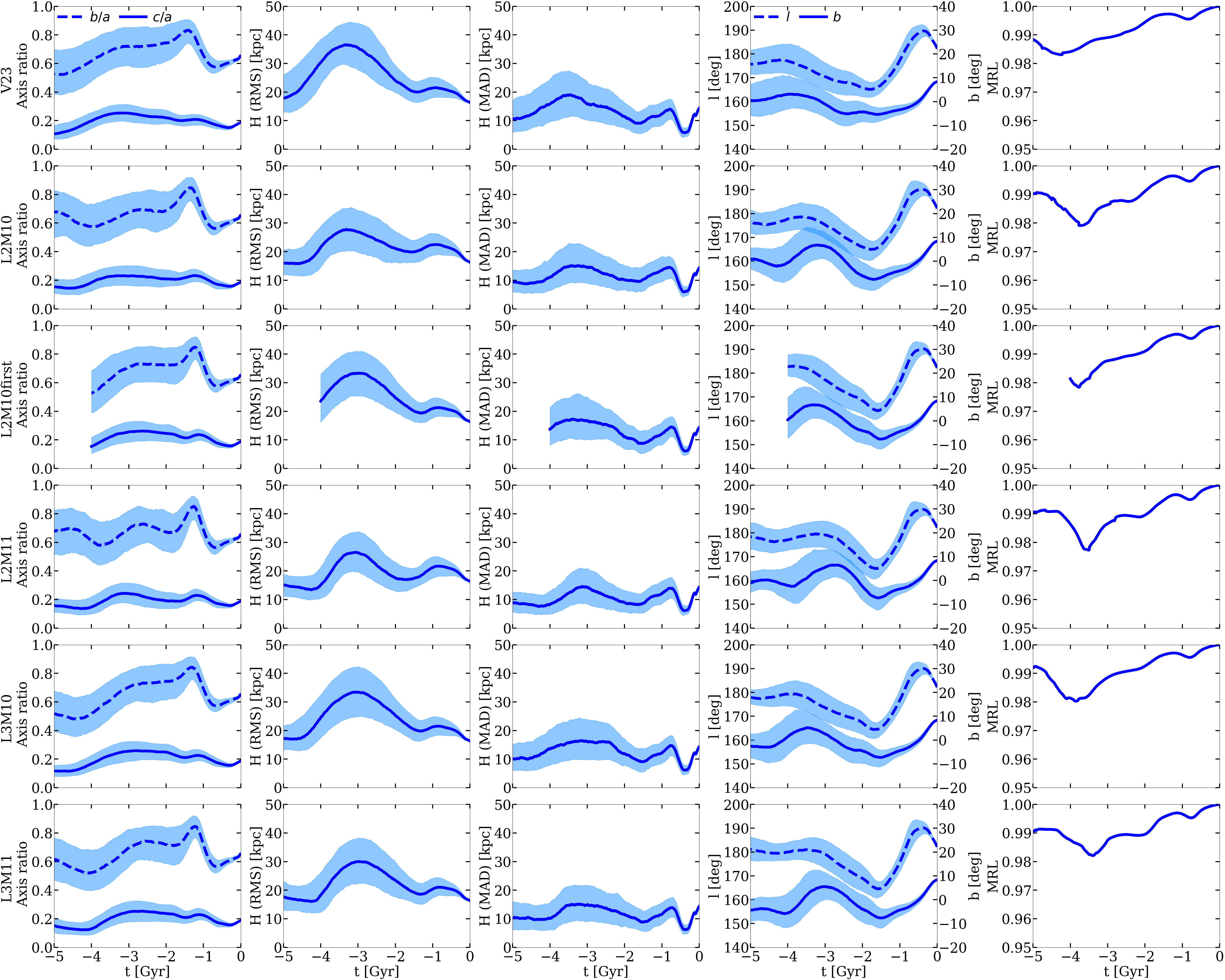}
    \caption{Time evolution of the structural parameters of the VPOS measured from the sample of  on-plane MW satellites only. Each row of panels represents the parameters for a different gravitational potential. Columns of plots represent from left to right: (1) the short-to-long axis ratio, $c/a$ (solid line), and intermediate-to-long axis ratio, $b/a$ (dashed line), (2) the thickness of the VPOS measured with the RMS, (3) the thickness of the VPOS measured with the MAD, (4) the Galactic longitude ($l$, dashed line) and latitude ($b$, solid line) of the direction of the normal vector of the VPOS, and (5) the MRL of the normal vector, all of them as a function of time. Lines represent the median values of the different metrics, and shaded areas the 16th and 84th quantiles.}
    \label{fig:VPOSevolutionMW}
\end{figure*}

In order to assess the long-term stability of the VPOS, we integrated the orbits of the on-plane galaxies within six time-evolving MW–LMC gravitational potentials. This allowed us to examine how the VPOS’s thickness, shape, and orientation evolve over time, and to determine whether its present-day configuration is a transient alignment or a persistent feature.
To minimize biases in the characterization of the VPOS, we separated its members into two subsamples: one including only MW satellites, and another combining both the MW and LMC satellites. The LMC and its satellites form a compact, coherent group infalling from large distances. At early times, their spatial separation from the MW satellite population creates a bimodal distribution in the 3D positions of the on-plane galaxies, which can bias the outcome of a PCA-based plane fit. To avoid this, we first analyze the VPOS long-term evolution using only on-plane MW satellites. We then examine the full on-plane sample,  but restricting the analysis to times after the LMC crosses within 250 kpc of the MW, a distance comparable to the MW’s virial radius and to the apocentre of some MW satellites (e.g., Draco, \citealt{Battaglia2022}). This ensures that the LMC satellites are not heavily distorting the planar fit. This separation of the on-plane galaxies in two subsamples not only mitigates possible biases but also enables us to assess the influence of the LMC on the evolution of the VPOS.

\subsubsection{Evolution of the VPOS from the MW on-plane satellites}
\label{sec:VPOSmw}
We begin by analyzing the evolution of the VPOS using only the on-plane MW satellites. In Figure~\ref{fig:VPOSevolutionMW}, we present the VPOS axis ratios $b/a$ and $c/a$, thickness, measured using both the RMS and the MAD, the direction of the normal vector in Galactic coordinates and its MRL, as a function of time. These metrics are shown for each of the six potentials as lines, with shaded regions representing their associated uncertainties. 

\begin{figure}
    \centering
    \includegraphics[width=0.95\linewidth]{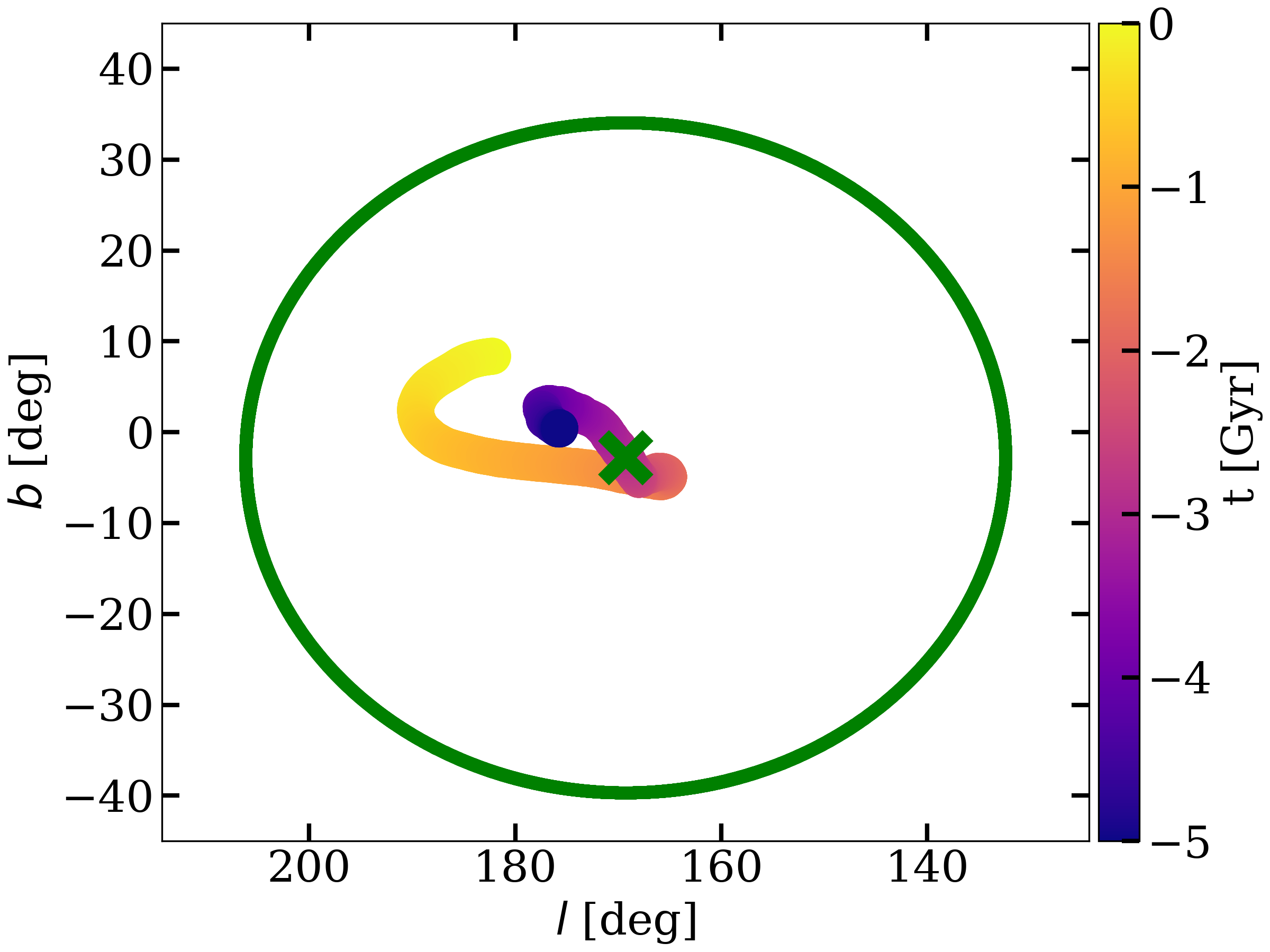}
    \caption{Evolution of the orientation of the VPOS. Color coded curve represents the time evolution of the direction of the normal vector of the VPOS in Galactic coordinates, measured from the on-plane MW satellites under the potential V23 (see Section~\ref{sec:VPOSmw}). Green cross represents the VPOS normal direction reported in \citet{PawlowskiKroupa2013} and green circle the border of the area around it encompassing 10\% of the sky.}
    \label{fig:normal_vector_orientation}
\end{figure}

The evolution of the axis ratios supports the long-term stability and planarity of the VPOS. Both $b/a$ and $c/a$ remain approximately constant across all gravitational potentials. In particular, $c/a$ is remarkably stable and consistently indicative of a flattened configuration. Its present-day values ($\sim0.2$) are in  agreement with previous VPOS characterizations (e.g., \citealt{Pawlowski2021}) and remain consistent throughout the last 5 Gyr. We note that for potentials V23, L3M10, and L3M11, the values of $b/a$ and $c/a$ tend to be slightly lower at early times compared to the other potentials. This is likely due to Sculptor coming from large distances under these models, which elongates the satellite distribution and thus lowers the axis ratios. However, we note that these differences are small. 
The VPOS thickness also remains stable over time regardless of the potential. At $t = 0$, both the RMS and MAD estimates yield a thickness of $\sim15$ kpc, in line with previous studies. The RMS-based thickness shows mild fluctuations, occasionally reaching up to $\sim35$ kpc, still within the range of thickness that current observational studies report. These variations likely reflect the complex orbital evolution of individual galaxies of the VPOS, particularly those with large apocentres, which can temporarily increase their projected distance to the plane, and thus increase its thickness. By contrast, the MAD measured thickness, being more robust to outliers, exhibits a remarkably flat trend, further backing that the VPOS has been a thin and persistent structure over time.
The orientation of the VPOS exhibits some temporal variations, likely driven by the orbital motions of its member galaxies. As observed with the thickness, on-plane galaxies that momentarily increase their distance to the plane can shift the orientation, leading to a temporary tilt.
To better illustrate the directional stability of the VPOS, in Figure~\ref{fig:normal_vector_orientation} we present an alternative visualization of the evolution of its orientation. The figure shows the Galactic longitude and latitude of the direction of the VPOS normal vector for potential V23 at each integration time, represented in a Cartesian 
$l$ vs. $b$ plot. While this is not an actual zoomed-in Galactic projection, it provides an approximation that allows us to better visualize directional changes over time. 
The normal vector remains well within the VPOS area throughout the entire integration time, indicating that, despite some fluctuations, the overall orientation of the structure is stable.  We note that results are very similar for the rest of potentials. The direction of the normal vector of the VPOS at $t=0 $ ($l = 182.2^\circ; b = 8.3^\circ$) differs from that reported by \citet{Pawlowski2013} in approximately 17 degrees, which is still well below the aperture defining the VPOS area ($\theta = 36.87^\circ$).
Such differences in the orientation of the VPOS between studies are likely attributable to the different galaxy samples used for its derivation.  Finally, the MRL indicates that the distribution of normal vectors across the MC iterations is highly concentrated, with values remaining above $\sim0.975$ throughout the whole integration time, highlighting the strong directional coherence of the structure.

Uncertainties in the metrics tend to grow with look-back time due to the propagation of observational errors through the orbit integration process. However, this growth is not monotonic; uncertainties typically stabilize beyond $t \sim -1.5$ to $-2$ Gyr. \citet{Kumar2025} investigated how observational uncertainties affect the inferred stability of satellite planes using  orbit integrations under MW-only potentials in a simulated VPOS. The study shows that while as $c/a$ may increase with time due to the  uncertainties, it eventually plateaus or stabilizes after a few Gyr. 
As also pointed out by \citet{Pawlowski2021b}, error sampling through a MC scheme, as done in this work, can be considered as applying uncertainties twice (first when measured, second when sampled), what can induce an artificial widening of the actual underlying  plane when measuring it and thus lead to an underestimation of its stability. Given the mean uncertainty in the PMs ($\sim 0.03$ mas yr$^{-1}$) and distance ($\sim4\%$) for the MW on-plane galaxies, and based on the results of \citet{Kumar2025}, we estimate that at $t = -3$ Gyrs, the measured $c / a$ ratio could  be  $\sim 60-95\%$ larger than that of the underlying true plane.
In our study, we do not find significant increase in $c/a$ over time, and the inferred thickness remains low and stable across all potentials. Even if some level of artificial thickening is present, our results show that the VPOS remains a thin and coherent structure throughout the 5 Gyr integration period, reinforcing the evidence of its long-term persistence and stability.

From $t = -2$ Gyr onward, we observe some subtle but potentially meaningful trends in the axis ratios and the direction of the normal vector. The $b/a$ ratio declines from $\sim$0.8 at $t \sim -1.5$ Gyr to $\sim$0.6 at $t \sim-$1 Gyr, suggesting an elongation of the satellite distribution within the plane, indicative of a more prolate configuration. Concurrently, the $c/a$ ratio exhibits a dip around $t \sim -0.5$ Gyr, indicating a transient vertical compression. This behavior is mirrored in the  MAD thickness metric, which also reaches a local minimum at that epoch.  Moreover, the orientation of the VPOS also shows changes over that time period. The reduced uncertainties at these recent times suggest that the changes are not a purely statistical fluctuation, but they are physically driven.
Indeed, the timing of these changes is particularly notable, as they coincide with the recent infall of the LMC into the MW’s inner halo. Given the LMC’s well-established gravitational influence on the MW satellite system (see \citealt{Vasiliev2023}), we interpret them as a result of the dynamical response of the MW satellites to the LMC’s proximity. In fact, we find that, over the past $\sim$1 Gyr, the distribution of the MW on-plane galaxies tends to elongate towards the position of the LMC. The orientation of the longest principal axis  measured for the distribution MW on-plane satellites also tends to align with the position of the LMC over the last $\sim$0.5 Gyrs.
However, we note that the dip in $c/a$ may arise from orbital phase effects: as most on-plane satellites are currently moving toward pericenter \citep{Taibi2024}, their past radial distribution would have been more extended, increasing the major axis while leaving the minor axis roughly unchanged.
The orientation of  the VPOS is particularly sensitive to spatial rearrangements of its member galaxies. Although the LMC’s orbital path is well aligned with the VPOS, differing by only $\sim5-15$ degrees prior to the LMC’s entry into the MW’s virial radius, even this modest offset induces subtle but measurable shifts in the VPOS normal vector. As a result of it, while the LMC approaches the VPOS from one side of the plane,  its gravitational influence on the MW satellites alters their distribution and thus the VPOS orientation, tilting it, increasing both $l$ and $b$. Once the LMC trespasses the plane and moves away from it, the MW on-plane galaxies tend to  rearrange accordingly, leading to the sudden change in the orientation of the VPOS at $t \sim -0.5$ Gyrs  (notice the decrease of $l$ from $t \sim 0.5$ Gyrs onwards in Fig.~\ref{fig:VPOSevolutionMW} or the U-turn in the VPOS direction in Fig.~\ref{fig:normal_vector_orientation}).
To sum up, the LMC induces mild changes in the structural properties of the VPOS, without significantly altering its overall configuration or compromising its stability. This is likely a consequence of the close alignment between the LMC’s orbital path and the VPOS, which allows the otherwise intense gravitational influence of the LMC on the MW satellites during its recent approach to have a relatively smooth and coherent effect on the on-plane satellite population. This raises important questions about the potential involvement of the LMC in the formation and evolution of  the VPOS.

Our results confirm that the VPOS has remained a stable structure throughout the entire integration time, regardless of the gravitational potential adopted for the orbit reconstruction. This allows us to place a lower limit on its age: the VPOS has persisted for at least 5 Gyr. Since our analysis so far considers only the MW satellites of the VPOS, we conclude that the planar structure was already in place around the MW well before the LMC crossed the virial radius. This rules out the possibility that the VPOS was formed as a result of the LMC’s recent infall.

\subsubsection{Evolution of the VPOS from the MW and LMC on-plane satellites}
\begin{figure*}
    \centering
    \includegraphics[width=0.95\linewidth]{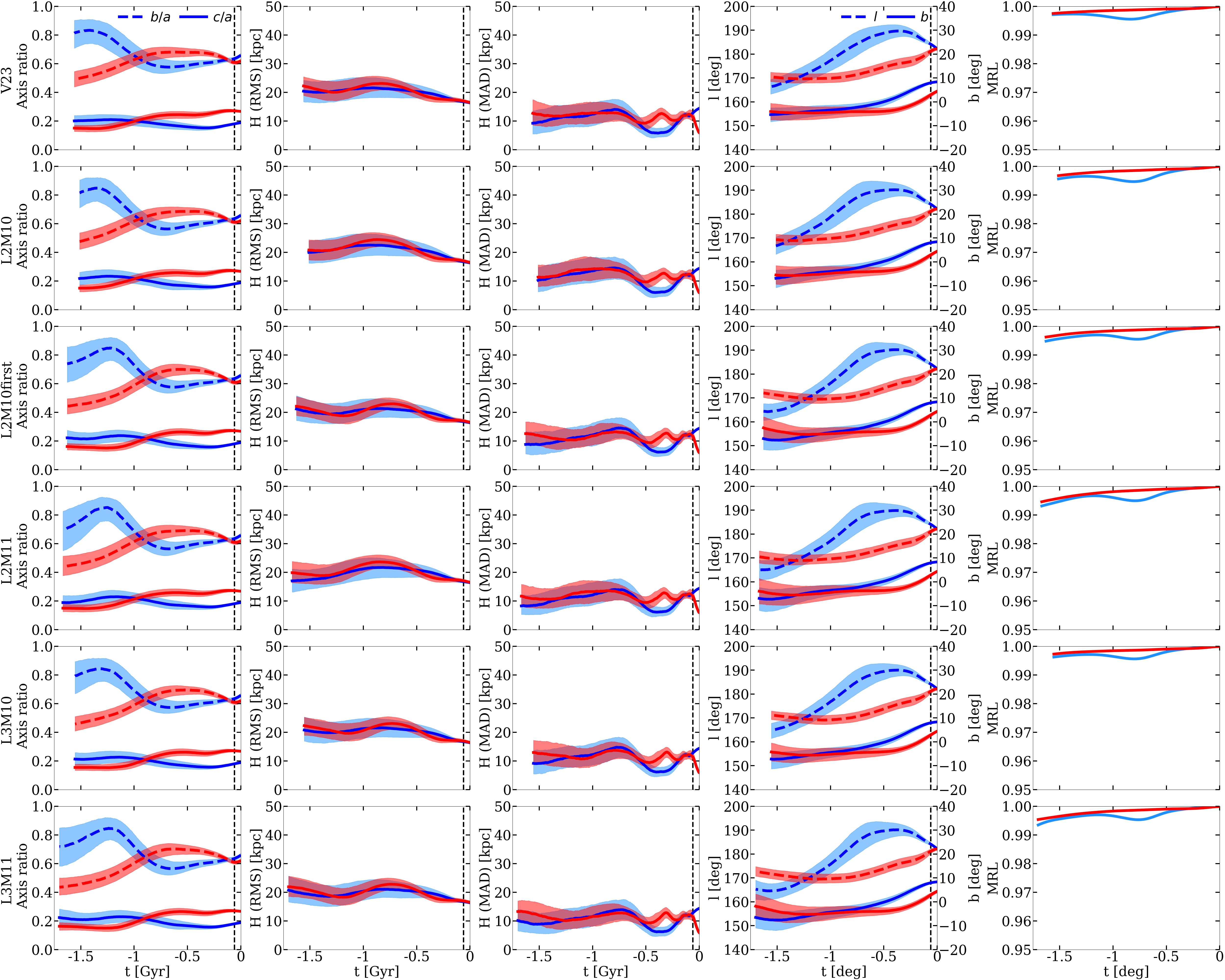}
    \caption{Time evolution of the structural parameters of the VPOS measured from the MW on-plane satellites only sample, and the MW and LMC on-plane satellites sample. The distribution of plots and markers coincides with Fig.~\ref{fig:VPOSevolutionMW}, however the analysed time span is shorter, starting when the LMC  gets within 250 kpc from the MW (see Section~\ref{sec:vposevo}). Red lines and shaded areas represent the parameters of the VPOS measured with the MW+LMC on-plane satellite sample, blues one are for the MW sample only. Dashed vertical lines show the time at which the LMC performed its pericentre.}
    \label{fig:VPOSevolutionMWLMC}
\end{figure*}

We now consider the evolution of the VPOS including the on-plane LMC satellites in the analysis. Figure~\ref{fig:VPOSevolutionMWLMC} shows the same  metrics as Figure~\ref{fig:VPOSevolutionMW}, but evaluated from the moment the LMC crosses within 250 kpc of the MW. For comparison, the figure  also includes the results for the MW-only sample.

The structural properties of the VPOS when measured for the whole on-plane sample are consistent for the six potentials. 
The axis ratios indicate that the VPOS also maintains a stable, planar configuration when LMC satellites are considered. At early times ($t \sim -1.7$ Gyr), the full satellite distribution appears more elongated and flatter, as seen from lower $b/a$ and $c/a$ values. This is because the distant LMC system still makes the galaxy distribution  bimodal and biases the PCA analysis. As the LMC and its satellites approach the MW and  blend with the MW on-plane population, both ratios rise and eventually stabilize, reaching values similar to those obtained with MW-only sample. Once such bias vanishes, it can be seen how the overall evolution of the ratios confirms that the VPOS remains a flattened, stable structure, regardless of the sample considered.
The thickness of the VPOS also shows a consistent and stable trend across both samples. 
Unlike the axis ratios at early times, the thickness is very similar whether or not the LMC satellites are considered in the sample, indicating that the VPOS thickness closely matches the spatial extent of the LMC satellite system, and that both the orbit of the LMC and the VPOS are well aligned. Notably, the MAD-based thickness dips slightly near the LMC’s pericentre, suggesting a temporary tightening of the structure, likely driven by the gravitational interaction of the LMC satellites with the MW at that time.
Regarding the orientation of the VPOS, we note that its evolution is broadly comparable for both samples, showing similar values at both the beginning and end of the analyzed time interval. The inclusion of the LMC satellites in the characterization of the VPOS reduces the variations of the orientation observed in the MW on-plane sample, showing a more stable configuration. The values of the MRL are very similar between both samples. 

Overall, we find that the structural metrics of the VPOS are very similar between the MW on-plane satellites, and the MW and LMC satellite samples once the LMC is sufficiently close to the MW to no longer bias the PCA analysis. The LMC and its satellites blend effortlessly with the MW on-plane galaxies, as they approach the MW, leaving the VPOS nearly unchanged. 
In particular, at present time, the thickness, shape, and orientation of the plane are remarkably consistent, regardless of whether the LMC satellites are considered.

\subsection{Stability of the VPOS in the literature}
\label{sec:lit}
Several observational studies have tried to assess the persistence of the VPOS using orbit integration, and unanimously conclude that it is a transient phenomenon (\citealt{Lipnicky2017, Maji2017, Sawala2023}). Such analysis are focused on the 11 'classical' MW satellites (namely: Carina, Draco, Fornax, Leo I, Leo II, LMC, SMC, Sagittarius, Sculptor, Sextans, and Ursa Minor) and  perform orbit integration under static MW-only potentials. \citet{Lipnicky2017} performed backward integration using PMs from \textit{HST}, finding that the VPOS goes from thickness 21.3 kpc at $t = 0$ to disperse within 1 Gyr. They performed tests using several subsets of the classical satellites (whole sample, not considering Leo I and Leo II, and only with the best measured PMs), finding always the same fate for the VPOS. \citet{Maji2017} forward integrated the orbits of the classical dwarfs for a Gyr under 2 different MW potentials, finding that the VPOS members are going to abandon the planar structure. Their planar metrics  $c/a$ and the thickness go from 0.18 and 19.6 kpc to 0.35 and 45 kpc at $t =  -0.5$ Gyrs and finally 0.42 and 64 kpc at $t = -1$ Gyr. 
More recently \citet{Sawala2023}, used PMs from Gaia EDR3 to forward and backward integrate the orbits of the classic satellites, finding that $c/a$ increases from $\sim 0.2$ to $\sim 0.3$ when advancing 0.5 Gyrs towards the future or 0.3 towards the past, concluding that the VPOS is transient. 

Our results show important differences with these studies. Although we obtain similar values of $c/a$ and thickness at present time, we do not find evidence of the VPOS vanishing, but the opposite. 
One possible explanation for these discrepancies is that previous studies are based  on MW-only gravitational potentials for the orbit integration. However, the significant influence of the LMC on the orbits of MW satellites has been repeatedly demonstrated in recent years (\citealt{Vasiliev2023}) and it is therefore essential to account for its effects when studying the stability of the VPOS. Gravitational potentials that include both the MW and the LMC provide a more realistic framework for capturing the complex orbital evolution of the galaxies in the vicinity of the MW. We note that the different modeling of the MW potential between previous studies and this work can also have an impact on the orbit reconstruction, nevertheless the inclusion of the effect of the LMC is likely to overshadow such differences.
However, the most likely source of disagreement lies in the galaxy samples used to characterize the VPOS. The commonly studied set of 11 classical MW dwarf satellites presents some possible issues. While our analysis shows 7 of them to be clearly on-plane, Sagittarius and Sextans  lie off-plane, and Leo I and Leo II are uncertain cases. Leo I and Leo II, despite being classified as 'uncertain', can potentially modify the outcome of the analysis; not only are they located at large distances from the  MW ($> 200$ kpc, \citealt{McConnachie2012}), but also their ambiguous membership to the VPOS  is a result of a  considerable scatter in their phase-space coordinates, having both galaxies large uncertainties in their PMs and distances. The conjunction of these two factors, plus the inclusion of two off-plane galaxies in the analysis, can severely bias any stability or persistence assessment.

In order to understand the origin of the discrepancies  with previous studies, we repeated our analysis of the VPOS stability using as galaxy sample the classical satellites of the MW (see Appendix~\ref{sec:app1}). In Figure~\ref{fig:VPOSevolutionMWclassic} we present the temporal evolution of key structural metrics of the plane defined by this subset of 11 galaxies. The axis ratios and thickness at $t=0$ support the presence of a planar configuration; however, this structure rapidly widens. The thickness increases quickly, particularly in the RMS-based measurement. The MAD-based thickness remains comparatively lower, reflecting its robustness to outliers and its better tracking of the stable, on-plane galaxies of the sample, though it also exhibits fluctuations.
The orientation of the fitted plane also evolves substantially, deviating by more than 60 degrees from its initial direction and temporarily departing from the VPOS region before returning. In parallel, the MRL decreases with time, indicating that the normal vectors become increasingly dispersed across MC iterations. 
We conclude that the fast thickening and erratic orientation of the plane is driven by the off-plane and uncertain VPOS members, which introduce a significant scatter in the galaxy distribution and thus have a significant impact on the PCA analysis. We note that Leo I and Leo II are the main contributors to the rapid widening of the plane, given that their large uncertainties in PMs and distances lead to poorly constrained orbits that strongly influence the plane fit.
The presence of on-plane galaxies in the sample of the classic satellites helps mitigate this effect to some extent, but it is not enough to  keep the stability of the structure. Therefore, we conclude that the differences with previous studies regarding the stability and persistance of the VPOS are most likely due to the sample of galaxies used. 
 This highlights the importance of carefully selecting the VPOS members, as including potential outliers has a dramatic impact on the study of its stability.

\subsection{Origin and evolution of the VPOS}
The analysis presented in this work shows that the VPOS is not a chance alignment in the present-day distribution of some dwarf galaxies around the MW. Understanding how this planar arrangement of  galaxies formed and, more importantly, persisted over time could provide valuable insights into the  history of the MW system and also help us understand how planes of satellites have formed elsewhere. 
The origin of the VPOS is a highly debated topic with no definitive solution yet (\citealt{Pawlowski2018}). However, over the last years, several studies have  helped to constrain the range of possible formation scenarios.

One of the proposed scenarios to explain the origin the VPOS is the preferential accretion of matter through filaments of the cosmic web, which would result in preferred directions for the accretion of satellites (\citealt{Zentner2005, Libeskind2005, Libeskind2011, Libeskind2014, Lovell2011}). One fundamental problem of this scenario is that the size of filaments that feed MW/M31-like galaxies are larger that their viral radii (\citealt{VeraCiro2011}), thus they are  unlikely to give rise to planar structures as thin as the VPOS. Additionally, studies based on cosmological simulations have shown that although satellites accreted through filaments can cause some anisotropy, it is not enough to reproduce the VPOS (\citealt{Pawlowski2012b}).

Another scenario that has been extensively explored is the tidal dwarf galaxy (TDG) hypothesis, in which the VPOS would be formed from these second-generation galaxies created during a past major merger or close encounter between the MW and another galaxy (e.g. \citealt{Fouquet2012, Akib2025}). In this framework, TDGs arise from the tidal debris of such an interaction, forming along a common tidal tail and thus naturally sharing a common orbital plane and direction of motion \citep{Pawlowski2011}. However, a significant challenge for this scenario lies in the expected absence of dark matter in TDGs. As these systems are not embedded within dark matter halos, they would be expected to exhibit low mass-to-light ratios, an expectation that is inconsistent with observational evidence for MW satellites, which typically show mass-to-light ratios $\geq 10$ M$_{\odot}$/L$_{\odot}$  (e.g., \citealt{McConnachie2012}). However, these problems can be bypassed by  adopting a series of alternative approaches like recurring cosmological frameworks different from $\Lambda$CDM (\citealt{Bilek2021}), among others (see \citealt{Pawlowski2018, Pawlowski2021}). One advantage of this scenario is that it can be directly tested by studying the observational properties of the on-plane and off-plane galaxies, which are expected to be  different given their separated origin. \citet{Taibi2024} 
did so by examining the properties of the galaxies of the MW-system, not
finding a significant preference for on-plane and off-plane dwarfs to follow different scaling relations, more specifically  by analysing their values on the luminosity-metallicity relation. 
Similar results have been reported for the plane around M31 (\citealt{Collins2015}). These results reduce the likelihood of the planar structures of the MW and M31 to be explained by TDGs.

Group infall has been proposed as another potential solution to explain the formation of planes of satellites (e.g. \citealt{LiHelmi2008, Donghia2008, Jerjen2025}). The accretion of a group of dwarfs would result in the dispersion of its galaxies along a common plane around the host. Once the galaxies are accreted they  would  still share similar direction, energy and angular momentum. Recently, it has also been explored the role of group infall in shaping the MW stellar halo (\citealt{CallinghamHelmi2025}), and potential evidence of a group accretion event has been reported among the MW satellites (Leo II, Leo IV, and Crater 1 , \citealt{Julio2024}). Also recently,
\citet{Taibi2024} studied the orbital histories of the MW on-plane systems finding that the co-orbiting ones are all currently moving towards the pericentres of their orbits, and had performed their apocentes $\sim 1$ Gyr ago. This striking similarity in the orbital phases of the on-plane MW satellites suggests that they may have originated from a group accretion event, in which multiple galaxies entered the MW halo together relatively recently.

The group infall hypothesis is particularly compelling in light of the ongoing accretion of the LMC system onto the MW. Having recently completed a pericentric passage, the LMC has significantly influenced the MW and its satellites throughout its trajectory across the Galactic halo (see \citealt{Vasiliev2023} for a review), and has also been proposed as a possible culprit of the formation of the VPOS. One proposed scenario is that the LMC could induce the clustering of the orbital poles of MW satellites (\citealt{GaravitoCamargo2021}); however, later studies argue that this effect alone cannot account for the formation of the VPOS (\citealt{CorreaMagnus2022, Pawlowski2022}). Our analysis agrees that the recent passage of the LMC is not responsible for the formation of the VPOS, however we find that it reoriented the orbital pole of Grus II, bringing it into alignment with the structure.  An alternative scenario posits that if the LMC is on its second pericentric passage, the VPOS could have naturally emerged from the stripping of some LMC satellites during the first pericentre (\citealt{Vasiliev2024}). 
Previous works had also proposed a possible association between some of the on-plane galaxies and the LMC. \citet{Lynden-Bell1976} suggested that the galaxies known at the time to lie in the plane could have been tidally stripped from the “Great Magellanic Galaxy,” the alleged progenitor of the LMC. Some subsequent studies similarly argued that several of the classical MW satellites may have once been associated with the LMC (\citealt{Nichols2011}) and some remarked the close connection between the VPOS, the LMC, and the Magellanic Stream (\citealt{Pawlowski2013, Pawlowski2015}).
Although our study focuses on the stability of the VPOS and is agnostic regarding its origin, our findings consistently point to a possible link between the formation of the VPOS and the early orbital history of the LMC. We now explore the evidence and implications of such connection.

\subsection{The LMC and the origin of the VPOS}
First, we note that the orbital path of the LMC aligns well with the VPOS (see Fig.~\ref{fig:plane_orbits}, and Section~\ref{sec:VPOSmw}). If the LMC is indeed on its first infall into the MW halo (\citealt{Besla2007}), such chance alignment would be highly improbable (\citealt{Metz2008}). The odds of an external group arriving from large distances along a trajectory that precisely traverses a long-lasting planar structure around the MW are very low, assuming they are completely unrelated. This apparent coincidence becomes even more surprising if we consider that the direction of motion of the LMC system around the MW also aligns with the sense of rotation of the VPOS MW satellites. To quantify this, we estimated the probability that two randomly oriented vectors are separated by an angle less than $\theta$, taken here to be between $\sim5^\circ$ and $15^\circ$, consistent with the offset between the LMC orbit and the VPOS  prior to the LMC entering the MW's virial radius. This corresponds to the solid angle of a spherical cap of angular radius $\theta$, normalized by the total surface area of a sphere, yielding a probability of approximately $\sim0.2$–$1.7\%$. Including the additional condition that the orbital sense must also match further reduces this probability by half. Altogether, this suggests that the LMC and its satellites are not merely recent additions to the planar structure around the MW.

Secondly, not only the LMC orbital path is well aligned with the VPOS, but also its passage by the MW does not  disrupt or significantly alter the structure. The LMC is known to exert a very significant gravitational influence in the MW satellites, streams and the MW itself (see \citealt{Vasiliev2023} for a review). Although we observe some effects of such influence on the VPOS, the overall structural parameters remain largely unchanged. 
Moreover, once the LMC and its satellites are close to the MW they integrate into the plane without introducing drastic changes. All this suggests that the LMC system is not simply overlaid on the VPOS but may be fundamentally connected to its origin. 
Interestingly, it has been proposed in recent years that some of the brightest satellites of the MW may have  been associated with the LMC in the past. For instance, studies based on orbital integration suggest that there is a small possibility that it could  be the case of Carina and Fornax (\citealt{Patel2020, Battaglia2022}). However, \citet{Vasiliev2024} highlights important limitations of orbit integration for tracing such associations, particularly since uncertainties in phase-space coordinates may bias against the association with the LMC. The spread that introduce the uncertainties is more likely to separate the galaxies from the LMC than associate them to it. 
Instead, \citet{Vasiliev2024} employed N-body simulations of the interacting MW-LMC system to study the possible origins of observed MW satellites under the scenario where the LMC had performed two pericentres around the MW. The study shows that the probabilities of association with the LMC for the co-rotating on-plane MW satellites are not only non-negligible, but significant. In particular, for Carina, Draco, Fornax, and Ursa Minor, they are roughly 50\%, i.e. they are as likely to have come with the LMC  than  to originate elsewhere. In the case of Crater II, Grus II and Tucana IV, the probability is $\sim 25 \%$, lower, but still significant. While not definitive proof of the VPOS’s origin, these findings are nonetheless compelling, particularly when viewed in the context of the broader body of evidence.

Finally, we note that the thickness of the VPOS is very similar to the spatial extent of the LMC system.
We measured such extent at $t = 0$ using an iterative MC procedure consisting of $10^3$ realizations. In each iteration, the sky coordinates and distances of the LMC satellites were randomly sampled from normal distributions, and then transformed into Galactocentric coordinates. To allow for a consistent comparison with the VPOS thickness, the 3D positions of the LMC satellites were projected onto a plane centered on the LMC, defined such that its normal vector aligns with the LMC’s velocity vector. We computed the RMS and MAD of the projected distances of the satellites to the LMC. This procedure was repeated in each iteration, and the final estimates were obtained by averaging the RMS and MAD values across all realizations. We find that the median spatial extent of the LMC group is $28.0^{+0.9}_{-1.0}$ kpc based on the RMS and $10.5^{+0.9}_{-0.7}$ kpc based on the MAD.
It could be argued that the spatial extent of the LMC at $t = 0$ may be not representative,  since at present-time the LMC has just passed its pericentre and thus has been subjected to a very intense gravitational interaction recently. Therefore, we analogously calculated it for $t = -2$ Gyr, before it enters the MW halo. We measure similar values. For instance, under the potential L2M10 we obtain $35.9^{+4.4}_{+5.0}$ kpc and $9.6^{+3.6}_{-4.1}$ kpc measured for the RMS and MAD, respectively.  These values are well below the reported virial radius of the LMC ($\sim 75$–$150$ kpc; \citealt{Kacharov2024}) and are comparable to the typical median radial distance between LMC satellite analogues and their host found in cosmological simulations ($\sim 37$ kpc; \citealt{Santos-Santos2021}).
The extent of the LMC system closely matches the thickness of the VPOS. While this may explain why the VPOS thickness remains largely unchanged when the LMC system approaches the MW, it also carries deeper implications. A key test of the group infall hypothesis is whether observed dwarf galaxy associations are compact enough to form thin planes (\citealt{Pawlowski2018}). This has long been a challenge for the scenario, as most  dwarf associations are too extended, typically spanning $\sim 150$–200 kpc, to produce thin structures like the VPOS (\citealt{Metz2009}). However, this is not the case for the LMC-system and the VPOS, where the spatial extent of both systems are clearly consistent. This suggests that the LMC system could be able to give rise to a thin structure, like the VPOS.  Our results show that the LMC has not given rise to the VPOS during its recent passage, so considering all the evidence, it is possible that it was formed during a previous pericentre.

In the scenario where the LMC is currently having its second pericentric passage, and with that several of the on-plane MW satellites having a $\sim$50\% probability of past association with the LMC (\citealt{Vasiliev2024}), it is plausible that the VPOS originated from satellites stripped from the LMC during its first pericentre. Such a scenario naturally explains the observed features (close alignment with the LMC orbit, mild influence, and smooth integration of LMC satellites),
that otherwise appear as remarkable succession of very unlikely coincidences. 
However, while this scenario provides an elegant explanation for the origin of the VPOS, it remains possible that a number of the on-plane satellites  joined the structure through other means. The recent inclusion of Grus II in the VPOS, whose orbit appears to have been reoriented by the gravitational influence of the LMC, suggests that similar events could have occurred during a previous LMC pericentre, thereby aligning some additional MW satellites with the VPOS (a scenario analogous to the one explored in \citealt{GaravitoCamargo2021}).
Another interesting case is that of Sculptor. In several models (V23, L3M10, and L3M11), Sculptor appears to have arrived from large distances and to have performed a single pericentre around the MW recently, yet is a clear member of the VPOS. While it cannot be  ruled out that a limited number of satellites may align with the plane by chance, particularly given the size of the MW’s satellite population, Sculptor’s orbital history argues against a purely coincidental alignment. Our analysis shows that it came into close proximity with the LMC approximately $\sim$4 Gyr ago, within its tidal radius, suggesting that the LMC may have significantly influenced its current orbital configuration. 
Taken together, these considerations suggest that although the LMC may have primarily given rise to the VPOS during an early pericentre, some on-plane galaxies may be part of the VPOS due to a number of other phenomena, in which the LMC also seems to play a key role. 
We stress, however, that the origin of the VPOS is very complex question. In particular, the role of dwarf galaxy group accretion in formation planar satellite structures is still  not fully understood, and deserves further investigation \citep{Pawlowski2021}.

\section{Conclusions}
\label{sec:conclusions}
In this work, we have studied the long-term stability and persistence of the Vast Polar Structure (VPOS). We started by evaluating the membership to the VPOS of 58 dwarf galaxies in the vicinity of the MW. Using accurate 6D phase-space data, we backward integrated the orbits of the identified VPOS members over the last 5 Gyrs under six different time-evolving gravitational potentials that account for the mutual interaction between the MW and the LMC. This allowed us to reconstruct the past trajectories of these galaxies in order to assess the evolution of the VPOS.

We identify 15 galaxies that are currently part of the VPOS, including 5 classical and 4 UFD satellites of the MW, as well as 6 likely members of the LMC satellite system. While the overall population of VPOS member galaxies appears stable over time, we find that Grus II has likely joined the structure only recently, as a consequence of the reorientation of its orbit induced by the gravitational influence of the LMC. Through the analysis of the structural properties of the VPOS over time, we find that:

\begin{itemize} 
    \item The VPOS is not a transient arrangement of some dwarfs around the MW, but a long-lived and stable structure that has persisted over the last 5 Gyrs. The short-to-long axis ratio ($c/a \sim 0.2$), the  thickness ($\sim$ 15 kpc), and the orientation of the VPOS remain roughly  constant over time across all the tested potentials, and are consistent with previous observational estimates.
    
    \item The VPOS predates the LMC’s recent infall. 
    The MW satellites that are currently  part of the VPOS,  were already forming a planar structure well before the LMC crossed the MW’s virial radius. This rules out scenarios where the VPOS is a young phenomenon triggered by the recent passage of the LMC.

    \item While the LMC infall exerts a strong gravitational influence on the MW and its satellites, it only induces mild perturbations in the VPOS that do not compromise the plane’s structure or long-term stability.
    
    \item The structural properties of the VPOS remain consistent regardless of whether LMC satellites are included in the analysis. This indicates that, as the LMC group approached the MW, its satellites integrated smoothly with the existing MW on-plane population without significantly altering the plane’s overall configuration.

\end{itemize}

Our findings point to a close connection between the VPOS and the LMC, suggesting that the LMC’s current approach is not its first passage by the MW. The good alignment of the LMC’s orbit with the VPOS plane and the coherent motion of the MW satellites would be an unlikely coincidence if the LMC were on its first infall. Moreover, despite the LMC’s strong gravitational influence, the VPOS remains remarkably stable, and the arrival of LMC satellites leaves its structure largely unchanged, further hinting at a causal connection. The close similarity between the spatial extent of the LMC satellite system and the VPOS thickness supports the idea that it could have given rise to the VPOS, as forming such a thin plane requires a compact progenitor group. However, since our analysis shows that the VPOS predates the LMC’s recent infall, such a scenario would require the LMC to be at least on its second pericentric passage. Indeed, in such scenario, several MW satellites that are members of the VPOS have been shown to have a ~50\% probability of past association with the LMC (\citealt{Vasiliev2024}), they are as likely of having come with the LMC than to form elsewhere. Therefore, the VPOS would be the result of the stripping of some of the LMC satellites during its first pericentre. This second-infall scenario naturally accounts for the observed alignment, limited perturbation by the LMC potential, and smooth integration between the MW on plane satellites and the LMC's in the VPOS.
However, while this scenario can  plausibly explain the origin of the VPOS, some galaxies may have joined the structure in other ways. It is also possible that during an earlier passage, the LMC reoriented the orbits of some MW satellites, as appears to have occurred for Grus II recently.

Although our results suggest a strong connection between the VPOS and the LMC, we cannot directly trace the VPOS’s origin further back in time, as orbit integrations become unreliable beyond the timescales explored here. The role of group infall in assembling coherent planar structures like the VPOS remains  an open question, as does the detailed orbital history of the LMC system. Future \textit{Gaia} data releases, with improved PM measurements, combined with high-resolution cosmological simulations capable of resolving analogues of the LMC and the MW’s lowest-mass on-plane satellites, will be key to understanding the origin of the VPOS and the broader context of the formation of planes of satellites in the LG and beyond.

\begin{acknowledgements}
The authors thank the anonymous referee for the comments that
have helped to improve this paper.
We thank E. Vasiliev for making publicly available the code AGAMA (\citealt{AGAMA2018, AGAMA2019}), and for his assistance with questions regarding its use. 
The authors acknowledge financial support from the \emph{Severo Ochoa} grant CEX2021-001131-S (MICIU/AEI/10.13039/501100011033), the Spanish Ministry of Science and Innovation project PID2021-124918NB-C41 (MICIU/AEI/10.13039/501100011033, FEDER, EU), and the RyC-MAX grant 20245MAX008 (CSIC). A.~del Pino also acknowledges funding from the Ram\'on y Cajal fellowship RYC2022-038448-I (MICIU/AEI/10.13039/501100011033, co-funded by the European Social Fund Plus).
This work has made use of data from the European Space Agency (ESA) mission Gaia (https://www.cosmos.esa.int/gaia), processed by the Gaia Data Processing and Analysis Consortium (DPAC, https://www.cosmos.esa.int/web/gaia/dpac/consortium). Funding for the DPAC has been provided by national institutions, in particular the institutions participating in the Gaia Multilateral Agreement.
\end{acknowledgements}

\bibliographystyle{aa} 
\bibliography{biblio.bib}

\begin{appendix}
\section{Caveats}
\label{sec:app2}
Our work provides compelling evidence for the persistence and stability of the VPOS over the past 5 Gyr, and points to a possible physical connection between the plane and the LMC. However, several limitations related to the data and modeling must be acknowledged.

One intrinsic limitation of our study stems from the way we define VPOS membership: galaxies are classified based on the present-day direction of their orbital poles (see Section~\ref{sec:VPOSmem}). This method inherently favors the selection systems that are already moving along a similar plane and whose uncertainties are small enough for their poles to remain within the VPOS area in the majority of the MC realizations. Such a selection naturally increases the likelihood of recovering a stable, long-lived planar structure. Additionally, as noted in Section~\ref{sec:VPOSmem}, another limitation of this classification method is that it uses strict boundaries for the selection of on-plane galaxies, which can exclude potential VPOS members.

Our analysis incorporates all known sources of observational uncertainty regarding the phase-space coordinates of the analysed galaxies, including errors in distances, proper motions, and radial velocities. These uncertainties, along with those in the solar parameters used for the transformation to Galactocentric coordinates, were propagated through a MC scheme, in order to identify VPOS member galaxies and to backward integrate their orbits. This allowed us to trace a statistically robust ensemble of orbital histories, by exploring the full space of  parameters of the phase-space coordinates. As a result, our conclusions reflect the full range of plausible configurations of the VPOS supported by current observational data. We note that future data releases of the \textit{Gaia} mission will provide us with more accurate PMs, and thus will allow to further restrict the orbital histories of the VPOS member galaxies.  

In the orbit integration, satellites are treated as massless test particles in the time-dependent MW–LMC potential. This setup does not account for dynamical friction acting on the satellites themselves. While this omission could affect the reconstructed orbits of the most massive satellites (e.g. Fornax), we expect the impact on the VPOS analysis to be reduced, as the vast majority of satellites contributing to the structure are of sufficiently low mass that dynamical friction is negligible on the relevant timescales.

We note that the primary source of systematic uncertainty in our analysis stems  from our incomplete knowledge of the gravitational potentials of the MW and the LMC. Despite significant advances in the last years, these potentials remain poorly constrained (e.g. see~\citealt{Wang2020} for a review on the MW mass and potential). To address this, we employed six time-evolving potentials encompassing a realistic range of MW–LMC configurations based on current observational constraints. This allowed us to explore the stability and persistence of the VPOS under different MW-LMC configurations.
In addition to the potentials used in this work, we analysed the stability of the VPOS using some additional potentials based on V23. We first used a variation which keeps the same initial parametrization of the LMC but changes the DM halo of the MW (the baryonic component remains unchanged). In this case the MW halo follows truncated spherical NFW profile, with scale radius of 14 kpc and truncation radius of 300 kpc.  
We considered two additional potentials based in V23 and the spherical MW variation mentioned above, but considering a lighter LMC, with initial mass of $5\times10^{10}$ M$_{\odot}$, scale radius of 5.61 kpc and truncation radius of 56.1 kpc (see \citealt{Vasiliev2023} for a more detailed explanation of these potentials). Under these additional potentials, we consistently find that the VPOS has remained a coherent and stable structure over at least the past 5 Gyr.

Our conclusions are also limited by the current MW satellite census. Incompleteness due to selection effects, especially for low surface brightness systems or objects obscured by the Galactic disk, could lead to future discoveries that lead to the revision of both the membership and the coherence of the VPOS.
Another key source of uncertainty is the poorly constrained orbital history of the LMC. Whether the LMC is currently on its first or second pericentric passage remains unclear, and this ambiguity limits our ability to fully understand its role in shaping the VPOS and its long-term evolution.

To conclude, we note that while our results are robust across multiple potential models, they are inherently model-dependent. Significant revisions to the mass, structure or evolution of the MW–LMC system could alter the inferred stability and origin of the VPOS.

\section{The MW classic satellites and the stability of the VPOS}
\label{sec:app1}

\begin{figure*}[htbp!]
    \centering
    \includegraphics[width=\textwidth]{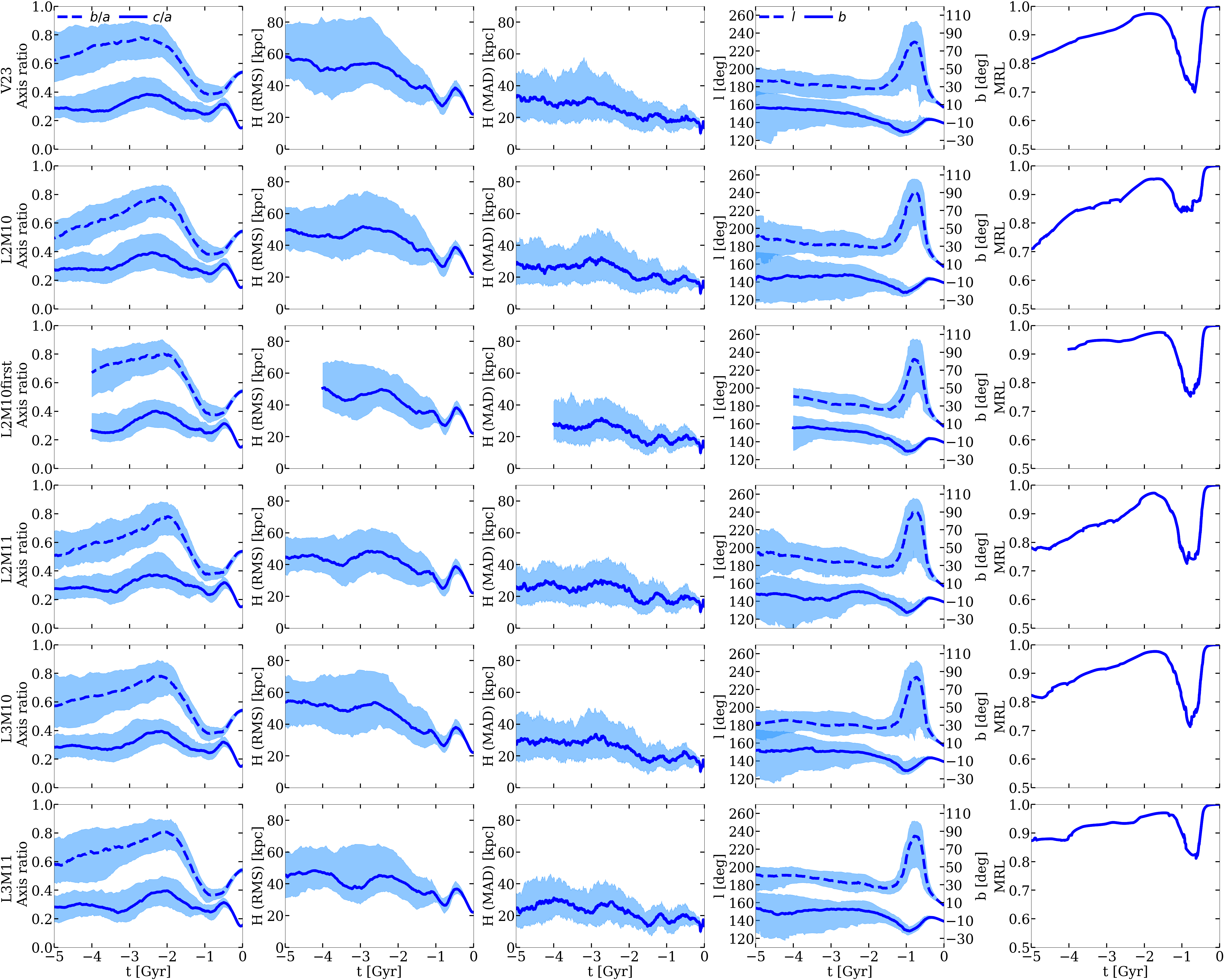}
    \caption{Time evolution of the structural parameters of the VPOS measured from the sample of the MW classic satellites. The distribution of plots and markers coincides with Fig.~\ref{fig:VPOSevolutionMW} , but the vertical axis ranges for thickness, orientation, and MRL are different}
    \label{fig:VPOSevolutionMWclassic}
\end{figure*}

To investigate the origin of discrepancies with previous studies,  we reanalyzed the stability of the VPOS using as sample the classical satellites of the MW, those used in earlier works (see Section~\ref{sec:lit}). Figure~\ref{fig:VPOSevolutionMWclassic} presents the time evolution of the structural properties of the planar configuration defined by this subset. This test provides insight into how the choice of galaxy sample influences the inferred stability of the VPOS.

\end{appendix}
\end{document}